\documentclass[aps,pra,preprint,superscriptaddress,longbibliography]{revtex4-1}

\usepackage{amsmath}
\usepackage{amsthm}
\usepackage{amsfonts}
\usepackage{amssymb}
\usepackage{xcolor,graphicx}
\usepackage{tikz}
\usepackage{color}
\usepackage[pdftex]{hyperref} 
\usepackage{longtable}
\usepackage{array}
\usepackage{dsfont}
	
\begin{document}
\title{Optimal crosstalk suppression in multicore fibers}
	
\author{B. Jaramillo \'Avila}
	\email[e-mail: ]{jaramillo@inaoep.mx}
	\affiliation{CONACYT-Instituto Nacional de Astrof\'{i}sica, \'{O}ptica y Electr\'{o}nica, Calle Luis Enrique Erro No. 1. Sta. Ma. Tonantzintla, Pue. C.P. 72840, Mexico.}	
\author{J. M. Torres}
	\affiliation{Instituto de F\'isica, Ben\'emerita Universidad Aut\'onoma de Puebla, Apdo. Postal J.48, Puebla, 72750, Mexico.}
\author{R. de J. Le\'on Montiel}
	\affiliation{Instituto de Ciencias Nucleares, Universidad Nacional Aut\'onoma de M\'exico, Apartado Postal 70-543, 04510 Cd. Mx., Mexico.}
\author{B. M. Rodr\'iguez-Lara}
	\affiliation{Tecnologico de Monterrey, Escuela de Ingenier\'ia y Ciencias, Ave. Eugenio Garza Sada 2501, Monterrey, N.L., Mexico, 64849.}
	\affiliation{Instituto Nacional de Astrof\'{\i}sica, \'Optica y Electr\'onica, Calle Luis Enrique Erro No. 1, Sta. Ma. Tonantzintla, Pue. CP 72840, Mexico.}

\begin{abstract}
We study propagation in a cyclic symmetric multicore fiber where the core radii randomly fluctuate along the propagation direction.
We propose a hybrid analytic-numerical method to optimize the amplitude and frequency of the fluctuations that suppress power transfer between outer and inner cores. 
Our predictions are confirmed by numerical experiments using finite difference beam propagation methods for realistic C-band fibers.
\end{abstract}
		
	\maketitle

Multi-core fibers with an underlying cyclic symmetry were a solution to the growing requirement for transmission capacity two decades ago. 
In original proposals, crosstalk was an issue to suppress in order to realize space division multiplexing while increasing the core density in these fibers\cite{Richardson2013,vanUden2014}.
Recently, the use of quasi-homogeneous structures proved a good approach to describe devices where asymmetry in parameters and materials suppresses crosstalk \cite{Takenaga2010,Takenaga2011}.

The idea of light localization in optical media, via transverse random fluctuations that are constant along the propagation axis, is related to Anderson localization of electron wave-functions in crystal lattices with static imperfections \cite{John1984,DeRaedt1989,Segev2013}. 
In optics and photonics this phenomenon is called transverse Anderson localization of light \cite{Mafi2015}.
In contrast, introducing fluctuations that also vary in the propagation direction produces faster than ballistic beam expansion, in a process similar to the so-called hyper-transport of light \cite{Levi2012} or environment-assisted quantum transport \cite{Rebentrost2009,Caruso2009,Leon2013,Leon2015,Guzman2016,Leija2018}.

A related avenue in experimental photonics aims to suppress crosstalk using homogeneous cores and inducing independent random fluctuations in core radii along the propagation direction \cite{Takenaga2010,Takenaga2011}.
These fluctuations are described by their statistical distribution, amplitude and frequency, that is, the number of variations per unit of length.
Our aim here is to optimize these parameters using hybrid analytic-numerical methods to gain insight of the underlying processes. 
In the following, we use coupled mode theory to optimize the maximum variation amplitude that produces feasible crosstalk suppression with an analytic model.
As an example, we use uniform random fluctuations which allow us to neglect all but the second moment of the random distribution.
Then, we numerically calculate the optimal variation frequency for these fluctuation amplitudes using coupled mode theory. 
Using the results from this approach, we construct a finite difference beam propagation model to produce a numerical experiment for feasible communication devices. 

\begin{figure}[!htbp]
	\begin{center}
		\includegraphics[scale=1]{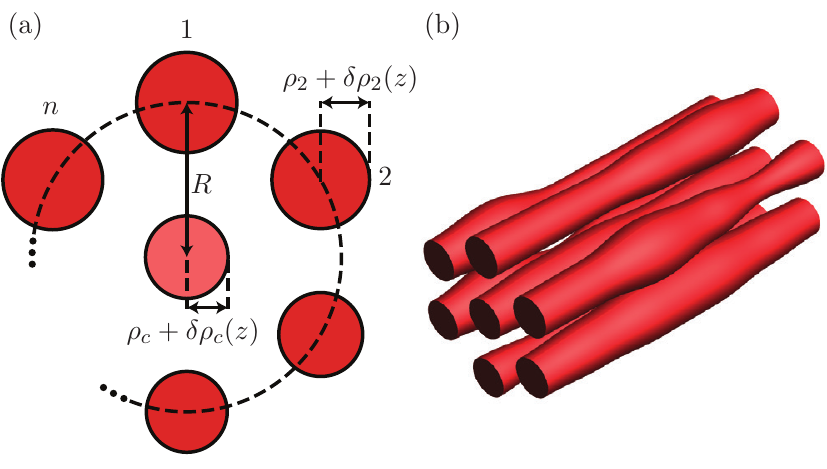}
	\end{center}
	\caption{(a) Cyclic symmetric multi-core fiber cross-section and (b) sketch for independent random variations in core radii.}\label{Fig:1}
\end{figure}

Our subject is a cyclic, multicore fiber composed by $n+1$ single-mode cores whose radii randomly varies along the propagation direction, $r_{j}(z) = \rho_{j} + \delta \rho_{j}(z)$ with $j=1,\ldots,n,c$. 
All external cores have the same reference radii, $\rho_{1} = \ldots = \rho_{n}$; the central core may have a different one, $\rho_{c}$, Fig.~\ref{Fig:1}. 
These fluctuations are inevitable during fabrication. 
They are usually kept well below the $2$\% mark and they can be introduced in a controlled manner above this threshold \cite{Takenaga2011}.
Small, smooth, and well-behaved variations induce slight deformations on the localized $\mathrm{LP}_{01}$ field modes at each core, as well as negligible back-propagation and power loss, and we can use coupled mode theory, 
\begin{align}\label{Eq:CoupledModeEquation}
-i \frac{\mathrm{d}}{\mathrm{d}z} \vec{\mathcal{E}}(z) = M(z) \cdot \vec{\mathcal{E}}(z),
\end{align}
to describe the propagation of just the complex mode amplitudes, 
$\vec{\mathcal{E}}(z) = \big( \mathcal{E}_{1}(z), \ldots,$ $\mathcal{E}_{n}(z), \mathcal{E}_{c}(z) \big)^{T}$.
The fluctuations induce variations in both the effective propagation and coupling constants, but the relative variation of the effective propagation constant is at least two orders of magnitude larger than that of the coupling constant.
Thus, we neglect the effect of these variations on the coupling constants,
\begin{align}\label{Eq:CoupledModeMatrix}
M(z)
=
\left(
\begin{array}{cccccc|c}
b_{1}	& g_{1}		& g_{2}		& \cdots	& g_{2}			& g_{1}		& g_{c}		\\
g_{1}	& b_{2}		& g_{1}		& g_{2}		& \cdots		& g_{2}		& g_{c}		\\
g_{2}	& g_{1}		& b_{3}		& g_{1}		& g_{2}			& \cdots	& g_{c}		\\
\vdots	& \ddots	& \ddots	& \ddots	& \ddots		& \ddots	& \vdots	\\
g_{2}	& \cdots	& g_{2}		& g_{1}		& b_{n-1}		& g_{1}		& g_{c}		\\
g_{1}	& g_{2}		& \cdots	& g_{2}		& g_{1}			& b_{n}		& g_{c}		\\
\hline
g_{c}	& g_{c}		& g_{c}		& \cdots	& g_{c}			& g_{c}		& b_{c}
\end{array}
\right),
\end{align}
with $b_{j} \equiv b_{j}(z)= \beta_{j} + \delta\beta_{j}(z)$, with $j=1,\ldots,n,c$. 
Here, all the external cores have the same reference propagation constant, $\beta_{1} = \cdots = \beta_{n}$, and the central core may have a different one, $\beta_{c}$. 
The coupling constants are independent of the propagation direction, $z$.
In the following, our numerics will refer to telecomm C-band, $1550$ nm, with standard silica fibers, $n_{c}=1.4479$ and $n_{cl} = 1.4440$.
The reference radii and center-to-center separation are $ r_{1,\ldots,n,c} = 4.5$ $\mu$m and $R=15$ $\mu$m, in that order. 
They yield approximated single-mode propagation constant $\beta = 5.859\,75 \times 10^{6} $ rad/m and intercore couplings $g = 256.636 $ rad/m. 
This model also describes laser inscribed waveguides where variations in the propagation constants are related to the radii and refractive index of individual waveguides, which can be controlled by the spot size and writing speed of the system.
Reported controlled random radii variations are in the 5\% range for multicore fibers \cite{Takenaga2010,Takenaga2011} and for femtosecond laser written photonic circuits the half-maximum reported variation in the refraction index is $3 \times 10^{-4}$ \cite{Blomer2006}.

The crux of our approach lies on two fundamental assumptions arising from the small, smooth, and well-behaved variations in the effective propagation constants. 
First, we model the fiber as a sequence of infinitesimal segments where the supermodes of each segment are provided by first-order perturbation theory on the supermodes of an homogeneous fiber.
Second, we treat propagation through the whole fiber as an averaging process for the initial impinging field independently propagating through each of these infinitesimal segments. 

In order to calculate the perturbed supermodes, we need to rewrite the coupled mode matrix,
\begin{align}
M(z) = M_{0} + M_{I}(z),
\end{align}
in terms of a constant matrix, $M_{0}$, and a diagonal perturbation matrix, 
$M_{1} = \mathrm{diag} \big( \delta \beta_{1}(z), \ldots,$ $\delta \beta_{n}(z), \delta \beta_{c}(z) \big)$.
The constant coupling matrix $M_{0}$ has a total of $n+1$ supermodes \cite{Jaramillo2019A}. 
Among these, $n-1$ supermodes are localized in the external cores and are given by 
\begin{align}\label{Eq:VectorModes:External}
\hat{S}_{j} &= \frac{1}{\sqrt{n}} \sum_{a=1}^{n} e^{-i \frac{2\pi}{n} j (a-1) } ~ \hat{e}_{a} \qquad \text{ for } j = 1, 2,\ldots, n-1,
\end{align}
where $\hat{e}_{j}$ is the $(n+1)$-dimensional vector with 1 in its $j$-th component and zero everywhere else. 
The corresponding $n-1$ propagation constants are
\begin{align}\label{Eq:PropagationConstants:External}
\lambda_{j} 
&= 
\beta \! + \!
\left\{
	\begin{array}{ll}
		\!\!\!
		2 \sum\limits_{k=1}^{m-1} \! \left\{ g_{k} \cos\left[ \frac{\pi}{m} j k \right] \right\} + g_{m}~(-1)^{j},
		& n \!=\! 2m, 
	\\
		\!\!\!
		2 \sum\limits_{k=1}^{m} \! \left\{ g_{k} \cos\left[ \frac{2\pi}{2m+1} j k \right] \right\},
		& n \!=\! 2m\!+\!1.
	\end{array}
\right.
\end{align}
The two additional supermodes are provided by 
\begin{subequations}\label{Eq:VectorModes:Central}
	\begin{align}
	\label{Eq:VectorModes:Central:1}
	\hat{S}_{n} &= -\sin\theta ~ \hat{S}_{0} + \cos\theta ~ \hat{e}_{n+1} ,
	\\
	\label{Eq:VectorModes:Central:2}
	\hat{S}_{n+1} &= \cos\theta ~ \hat{S}_{0} + \sin\theta ~ \hat{e}_{n+1},
	\end{align}
\end{subequations}
with propagation constants, 
\begin{subequations}\label{Eq:PropagationConstants:Central}
	\begin{align}
	\label{Eq:PropagationConstants:Central:1}
	\lambda_{n} &= \left( \lambda_0 + \beta_c - \sqrt{ \left( \lambda_0 - \beta_c \right)^{2} + 4 \, g_c^{2} \, n }, \right)/2, 
	\\
	\label{Eq:PropagationConstants:Central:2}
	\lambda_{n+1} &= \left( \lambda_0 + \beta_c + \sqrt{ \left( \lambda_0 - \beta_c \right)^{2} + 4 \, g_c^{2} \, n }, \right)/2.
	\end{align}
\end{subequations}
The difference between these constants yields the Rabi frequency
\begin{equation}
\Omega = \sqrt{ \left( \lambda_{0} - \beta_{c} \right)^{2} + 4 n ~ g_{c}^{2} }.
\end{equation}
Additionally, the mixing angle, 
\begin{equation}
\tan\theta = \frac{2 \sqrt{n} ~ g_{c}}{\lambda_{0} - \beta_{c} + \Omega},
\end{equation}
parametrizes the whole homogeneous fiber; each different realization of the multicore fiber can be described by its mixing angle.
In absence of coupling to the inner core, $g_{c} = 0$, these two supermodes are given by the central core mode, $\hat{S}_{n} |_{\theta=0} = \hat{e}_{n+1}$ with propagation constant $\beta_c$ and the mode $\hat{S}_{n+1}|_{\theta=0} = \hat{S}_{0}$ with propagation constant $\lambda_{0}$, where these are defined using Eq. (\ref{Eq:VectorModes:External}) and Eq. (\ref{Eq:PropagationConstants:External}), respectively, with $j=0$. 

\begin{figure}[!htbp]
	\begin{center}
		\includegraphics[scale=1]{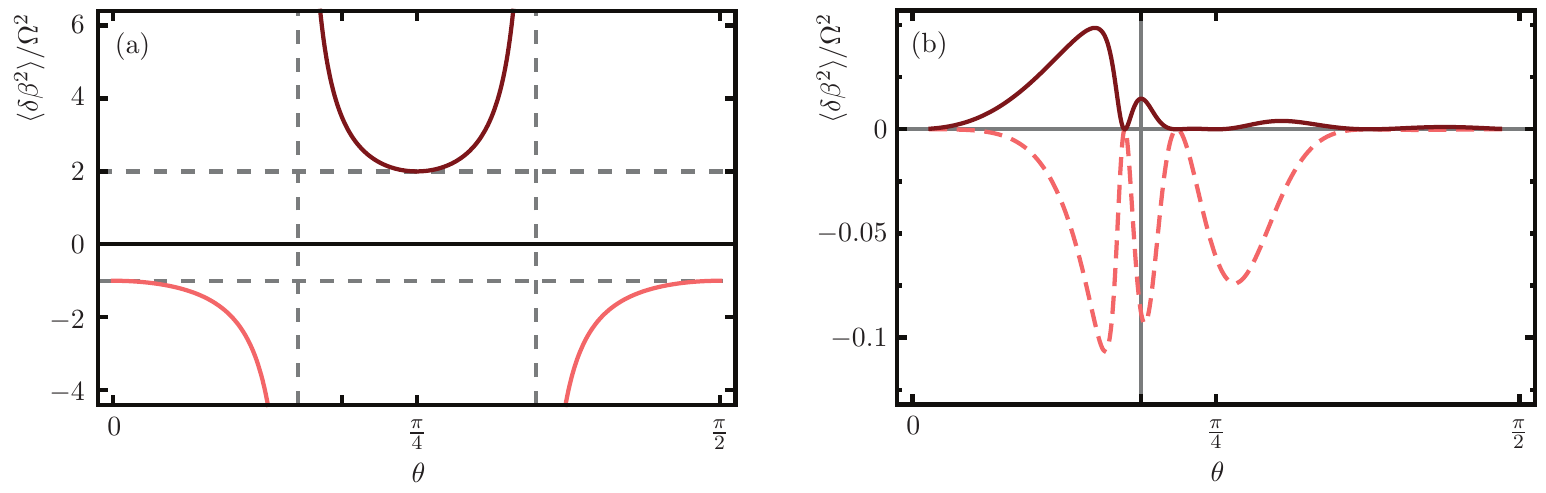}
	\end{center}
	\caption{Optimal fluctuation size divided by the Rabi frequency as a function of the mixing angle, $\theta$, for a (a) two- and (b) seven-core system. The vertical gray line marks the mixing core angle for our simulations.}\label{Fig:2}
\end{figure}

Now, we include the effects of the small propagation-dependent variations provided by $M_{I}(z)$. 
We use first-order perturbation theory to calculate the unnormalized supermodes for the infinitesimal segment at the propagation distance $z$,
\begin{equation}
\vec{A}_{j} = \hat{S}_{j} + \sum_{k=1,~k \neq j}^{n+1} \frac{ \hat{S}_{k}^{\dagger} \cdot M_{I} \cdot \hat{S}_{j} }{ \lambda_{j} - \lambda{k} },
\end{equation}
where $j=1,\ldots,n,n+1$. 
We aim for isolation between central and external cores after a given propagation distance. 
For this, we use a target state with equal field amplitude in the external cores and zero in the central core; 
the uncoupled supermode $\hat{S}_{0}$ is chosen as initial condition. 
We look for maximum overlap between our target state and the output after propagation through the fiber. 
This overlap is quantified by the inverse participation ratio (IPR) between the target mode and the propagation-dependent supermodes,
\begin{equation}
\textrm{IPR}[\hat{S}_{0},\hat{A}] = \sum_{j=1}^{n+1} \left| \hat{S}_{0}^{\dagger} \cdot \hat{A}_{j} \right|^{4},
\end{equation}
where $\hat{A}_{j}$ denotes the $j$-th normalized propagation-dependent supermode. 
A maximum IPR of one is reached when the target state overlaps with a single state of the basis and a minimum of $1/(n+1)$ when the target state overlaps with all states homogeneously.
It is worth noting that the IPR was introduced in the context of localization in disordered quantum system \cite{Wegner1980,Thouless1974}. 
Here, we employ it to quantify the closeness of a preferred mode to an eigenmode of the perturbed system where light localization is feasible. 
It is cumbersome to calculate the IPR of our target with the propagation-dependent basis as it involves integration over all infinitesimal segments. 
Instead, we argue that propagation through each infinitesimal segment will induce small changes on the initial field distribution.
Thus, instead of calculating the $z$-dependent propagation of an initial field distribution, we calculate the average of the output of that initial distribution through each infinitesimal segment.
This allows us to substitute any power of the local variations by the statistical moments of the corresponding power, $\delta \beta^{m} \rightarrow \langle \delta \beta^{m} \rangle$. 
For the sake of simplicity, we assume small random variations evenly distributed around zero and keep the leading non-vanishing moment $\langle \delta \beta^{2} \rangle$ to look for ideal crosstalk suppression, $\textrm{IPR}[\hat{S}_{0},\hat{A}] = 1$.
In principle, this leads to a relation between the second moment and the fiber parameters, that we can use to determine the optimal fluctuation amplitude that maximally suppresses crosstalk for a given fiber,
\begin{align}\label{Eq:OptimalNo}
&
	\frac{\langle \delta \beta^{2} \rangle}{n~\Omega^{2}} 
=
	\frac{ 1-\cos^{4}\theta-\sin^{4}\theta }{2}~ 
	\Big( 
		6 \cos^{4}\theta ~ \sin^{4}\theta 
		- \Omega^{2} ~ \sin^{6}\theta ~ \mathcal{S}_{n}
\nonumber \\ 
&
\qquad
		- \cos^{2}\theta ~ \sin^{6}\theta
		- \Omega^{2} ~ \cos^{6}\theta ~ \mathcal{S}_{n+1}
		- \cos^{6}\theta ~ \sin^{2}\theta
	\Big)^{-1},
\end{align}
where the auxiliary functions are defined by the expression
\begin{equation}
\mathcal{S}_{j} = \sum_{k=1}^{n-1} \frac{1}{(\lambda_{j}-\lambda_{k})^{2}} \qquad \text{ for } j = n,n+1.
\end{equation}
For example, random fluctuations with a uniform probability distribution in the range $[-\delta\beta_{\mathrm{max}},\delta\beta_{\mathrm{max}}]$ have a second moment $\langle \delta \beta^{2} \rangle = \delta \beta_{\mathrm{max}}^{2}/12$ and those with a Gaussian probability distribution have $\langle \delta \beta^{2} \rangle = \sigma^{2} ~ \delta \beta_{\mathrm{max}}^{2}$, where $\sigma$ gives the width of the Gaussian distribution and $\delta\beta_{\mathrm{max}}$ is given by the fluctuation size allowed by the experimental system.
Depending on the characteristics of the fiber, sometimes there is a feasible solution for maximal crosstalk suppression. 
Figure \ref{Fig:2}(a) shows the relation between the scaled second moment and the mixing angle for a two core system, $n=1$, where viable fluctuation amplitudes for maximum suppression are given by the positive branch of the equation. 
Figure \ref{Fig:2}(b) shows an example of unfeasible maximum suppression (dashed line) for a seven core system but where partial suppression (solid line), $\textrm{IPR}[\hat{S}_{0},\hat{A}] < 1$, is plausible. 

\begin{figure}[!htbp]
	\begin{center}
		\includegraphics[scale=1]{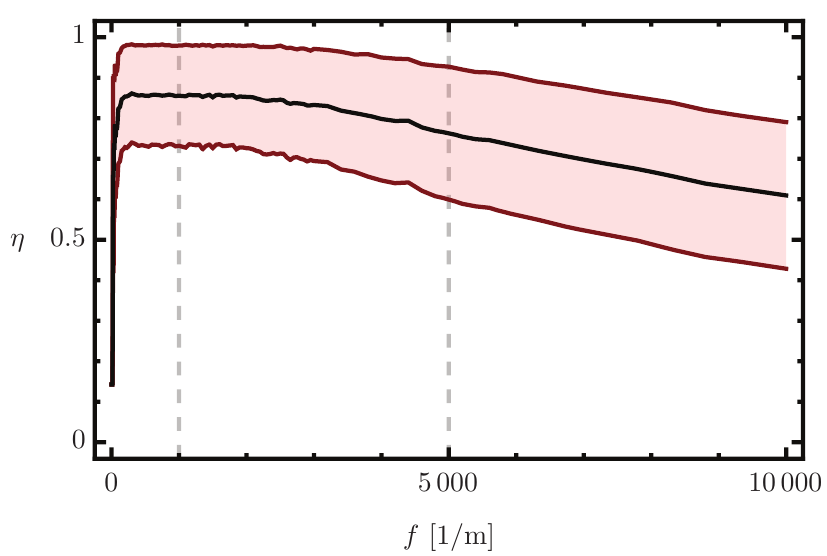}
	\end{center}
	\caption{Average and dispersion of irradiance localization in the external cores as a function of the number of random variations per unit length. The vertical gray lines mark the variation frequencies used in our simulations.}\label{Fig:3}
\end{figure}

Once the maximum feasible crosstalk suppression and corresponding fluctuation amplitude are assessed, we can use coupled mode theory to run a statistical numerical analysis to find the optimal number of variations per length unit that will produce the maximum available crosstalk suppression in a given propagation length, $z_{c}$. 
For the sake of simplicity, we focus on crosstalk suppression between the external and inner core. 
Our figure of merit will be the localization of the field in the external cores given by the ratio between the total irradiance in the external cores with respect to that in all cores, 
\begin{align}
\eta(z) = \frac{\sum_{j=1}^{n} \vert \mathcal{E}_{j}(z) \vert^{2}}{\sum_{j=c,1}^{n} \vert \mathcal{E}_{j}(z) \vert^{2}}.
\end{align}
Complete localization occurs for $\eta = 1$, complete transfer to the central core at $\eta = 0$, and complete delocalization in all cores for $\eta = n / (n+1)$.
Figure \ref{Fig:3} shows our rate of localization in the external cores for a seven-core fiber, $n=6$, with an uniform random fluctuation distribution with maximum effective propagation constant $\delta \beta_{\mathrm{max}} \approx 880$ rad/m corresponding to a radii variation of $8.5$\% or a refractive index variation of $3\times10^{-4}$. 
We chose as total propagation distance $z_{c} = 27 \times \pi / \Omega$; an odd integer multiple of the distance providing maximum power transfer to the central core.
Figure \ref{Fig:3} shows the average (solid black line) and the mean dispersion (light red area) of $5\,000$ independent realizations for up to $10\,000$ variations per meter using coupling mode theory.

\begin{figure}[!htbp]
	\begin{center}
		\includegraphics[scale=1]{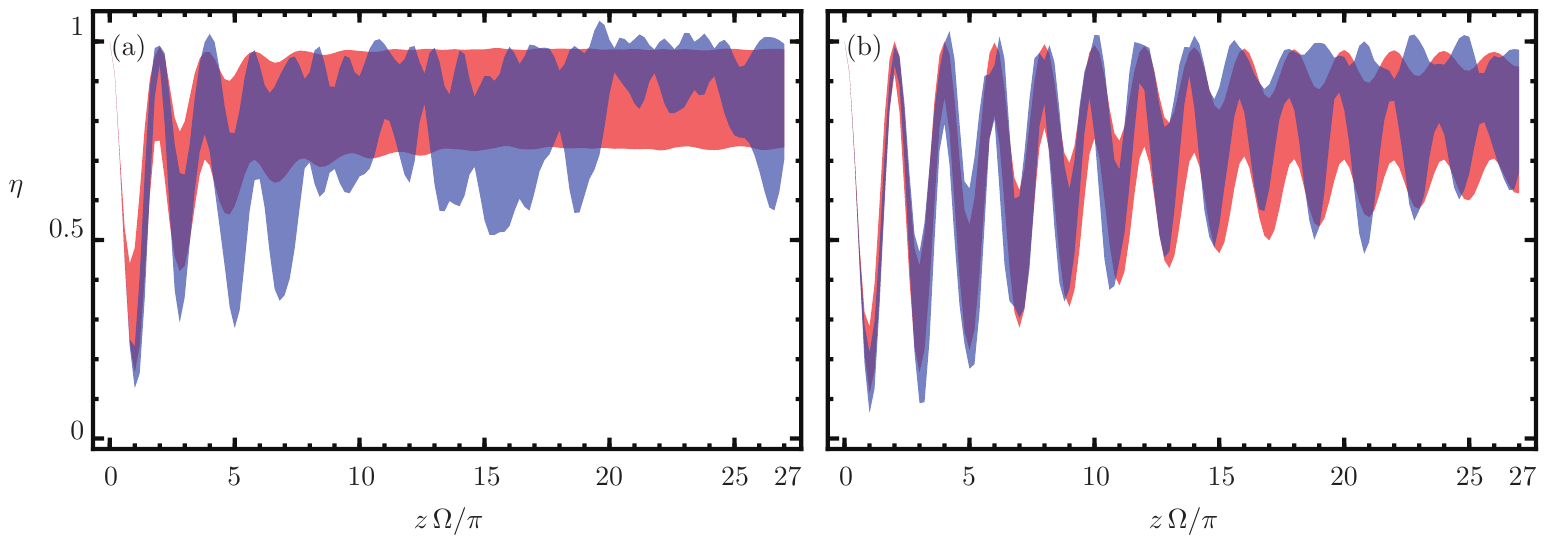}
	\end{center}
	\caption{Irradiance localization in the external cores from coupled mode (red region) and finite difference (blue region) propagation for (a) optimal and (b) sub-optimal variation frequency. The width of colored regions is two standard deviations centered around the average localization.}\label{Fig:4}
\end{figure}

Armed with our prediction, we use finite difference beam propagation methods to conduct a numerical experiment to try and confirm them.
Figure \ref{Fig:4}(a) and \ref{Fig:4}(b) shows the localization parameter versus the propagation distance in terms of the complete delocalization distance, $\pi/\Omega$, for $1\,000$ and $5\,000$ repetitions per meter, in that order, in the seven-core fiber of Fig. \ref{Fig:3}. 
The coupled-mode theory approach and the finite difference methods display slightly different delocalization distances, 
$\pi/\Omega = 2323.41~\mu$m and
$\pi/\Omega = 2155.80~\mu$m, respectively.
We show the average of $5\,000$ independent exact numerical realizations using coupled mode theory in red. 
In blue, we show the average of 5 independent realizations propagated using finite difference beam propagation methods. 
We want to emphasize that the $5\,000$ coupled mode and the $5$ finite difference realizations take an average of $3$ and $15$ hours to compute, in that order. 
Note that the numerically-predicted optimal number of random variations per unit length, Fig.~\ref{Fig:4}(a), stabilizes in a shorter propagation distance than the sub-optimal value, Fig.~\ref{Fig:4}(b), to similar values of the average irradiance localization. 

We have shown that assuming small, smooth, well-behaved, independent random variations in the core radii of a symmetric multicore fiber allows us to predict, in analytic form, the maximum variation amplitude that will produce crosstalk suppression between its supermodes. 
The analytic maximum variation amplitude, then, helps us numerically define an optimal number of variations per length unit that will produce a stable target suppression at a given propagation length. 
Our treatment is a simple, low computational resource, design method that is in good agreement with more resource intensive numerical methods and recent experimental results. 

\begin{acknowledgments}
B.J.A. acknowledges financial support by CONACYT under C\'atedra Grupal \#551. 
R.J.L.M. thankfully acknowledges financial support by CONACYT under project CB-2016-01/284372, and by DGAPA-UNAM under project UNAM-PAPIIT IA100718. 
B.M.R.L. acknowledges financial support by CONACYT under project CB-2015-01/255230, and by the Marcos Moshinsky Foundation under 2018 Marcos Moshinsky Young Researcher Chair. 
\end{acknowledgments}

%
	
\end{document}